# Is abundant A-to-I RNA editing primate-specific?

Eli Eisenberg[1,2,*], Sergey Nemzer[1], Yaron Kinar[1], Rotem Sorek[1], Gideon Rechavi[3] & Erez Y. Levanon[1,3].

[1] Compugen Ltd., 72 Pinchas Rosen St., Tel-Aviv 69512, Israel

[2] School of Physics and Astronomy, Raymond and Beverly Sackler Faculty of Exact Sciences, Tel Aviv University, Tel Aviv 69978, Israel

[3] Department of Pediatric Hematology-Oncology, Chaim Sheba Medical Center and Sackler School of Medicine, Tel Aviv University, Tel Aviv 52621, Israel

## Abstract

A-To-I RNA editing is common to all eukaryotes, associated with various neurological functions. Recently, A-to-I editing was found to occur abundantly in the human transcriptome. Here we show that the frequency of A-to-I editing in humans is at least an order of magnitude higher as that of mouse, rat, chicken or fly. The extraordinary frequency of RNA editing in human is explained by the dominance of the primate-specific Alu element in the human transcriptome, which increases the number of double-stranded RNA substrates.

A-to-I RNA editing is the site-specific modification of adenosine to inosine in stem-loop structures within precursor messenger RNAs, catalyzed by members of the double-stranded-RNA (dsRNA) specific ADAR (adenosine deaminase acting on RNA) family[1]. ADAR-mediated RNA editing is essential for the development and normal life of both invertebrates and vertebrates[2-5]. Altered editing patterns were associated with inflammation[6], epilepsy[7], depression[8], amyotrophic lateral sclerosis (ALS) [9] and malignant gliomas[10]. In a few known examples, editing changes an amino-acid in the translated protein, resulting in a change in its function. However, it was suggested that this might not the primary role of editing by ADARs[4], as most documented editing events occur within UTRs and intronic regions[11]. These editing events may affect splicing, RNA localization, RNA stability and translation[12], but full understanding of the purpose of editing in these regions is yet elusive.

Using a combination of bioinformatics to search for potential stem loop structures in transcripts combined with differences between EST and genomic sequences we have recently reported the identification of abundant A-to-I editing in human, affecting more than 1600 different genes[13]. Most of these editing sites reside in Alu elements within UTR regions[13,14]. Alu elements are short interspersed elements (SINEs), typically 300 nucleotides long, which account for >10% of the human genome. Despite being considered genetically functionless, Alu elements have been suggested to have broad evolutionary impacts[15,16]. They are found in all primates but in no other organism [17,18]. Therefore, they were suggested to play a role in the evolution of primates [19,20]. However, the nature of this role is still under debate. The question thus arises whether the abundance of A-to-I editing sites in humans is related to some special characteristics of the Alu repeat, and thus unique to primates, or whether similar editing patterns could also be observed in other organisms with a different, yet similar, composition of SINEs.

## Comparative search for RNA editing sites

In this study, we have searched for A-to-I editing sites in human, mouse (*Mus musculus*), rat (*Rattus norvegicus*), chicken (*Gallus gallus*) and fly (*Drosophila melanogaster*). We have found that the frequency of predicted A-to-I RNA editing in human is at least an order of magnitude higher than in other organisms. For this purpose, we used the algorithm described in Levanon et. al.[13]. Briefly, as ADARs bind to double-stranded RNAs (dsRNAs), the algorithm searches for potential dsRNA structures within the genomic sequence of each gene, and enumerates all deviations of the expressed sequence from the genomic one in these dsRNA regions. The sequencing reaction (as well as the ribosome) recognizes inosine as guanosine (G). Therefore, the fingerprints of ADAR editing are genomically encoded A's that are read as G's in the RNA sequence. A strict cleaning procedure is used to remove sequencing artefacts (noisy expressed sequence) and known SNPs, resulting in a very clean set of A-to-G mismatches, which, with high confidence, attest for A-to-I editing (see ref. 13 for details).

Application of this algorithm to the human transcriptome (more than 5 million ESTs and RNAs) has yielded 12,723 editing sites (with an estimated accuracy of >95%), in 1637 different genes[13]. A subset of the results has been experimentally validated, resulting in the observation of editing in 26 novel substrates and confirming the computationally estimated accuracy. Here, we applied this algorithm to the mouse transcriptome (over 4 million ESTs and mRNAs). Mouse compares to human in terms of the quality of the sequenced genome, genome size, number of genes, and the amounts of expressed sequences, making a comparison applicable. Using the exact same algorithm as for the human genome, we found only 302 A-to-I editing sites (estimated accuracy 90%), in 87 different mouse genes. The detected level of editing in mouse is thus 40-fold lower than that in human.

In addition, we have conducted another independent search for RNA editing, which does not use the above-mentioned algorithm, and is based on a different sequence alignment method. We used the UCSC alignments of human and mouse RNA sequences to their genome  ([http://genome.ucsc.edu](http://genome.ucsc.edu)) [21] and recorded all mismatches along them. 128,068 human RNA sequences (total length 259Mbp) and 102,895 mouse RNA sequences (total length 198Mbp) were scanned. The distribution of mismatches in these sequences is presented in Figure 1a and Table 1. Even a simple count of all mismatches exhibits a vast overrepresentation of A-to-G mismatches in human sequences, suggesting  that there are ~50,000 inosines in these sequences (~1:5,200bp). On the other hand, the number of A-to-G mismatches in excess of the noise background level in mouse RNAs is only ~3000 (~1:66,000bp), reflecting a 17-fold higher fraction of inosines human as compared to mouse. Again, the results point to a striking difference in editing patterns between human and mouse.

A-to-I editing sites often occur in clusters, an edited sequence typically being edited in many close-by sites[11] . We thus searched for sequences that exhibit three or more consecutive identical mismatches (see Figure 1d). A-to-G consecutive mismatches are vastly overrepresented compared to other types of consecutive mismatches in human, but not in mouse (Figures 1B and 1C). We found a set of 4864 human RNA sequences with three or more consecutive A-to-G mismatches, with estimated accuracy of 80%, suggesting ~4000 RNA sequences are actually multi-site edited (3% of the total number of RNA sequences). In comparison, the same analysis applied to mouse RNA sequences yielded an estimate of only ~220 (0.2%) multi-edited mouse RNA sequences (see Table 1). Here too, the number of edited sequences found in human is 20-fold higher than in the mouse. We thus conclude that A-to-I editing levels in human are at least an order of magnitude higher than in mouse. Note that after the submission of this work, similar results have been published by Kim et al[14].

To check whether the detected differences are primate or rodent specific, we repeated the above two analyses (single and multi mismatch counts in RNAs relative to genome) on the genomes of rat (10,999 RNA sequences), chicken (19,218 RNA sequences) and fly(14,632 RNA sequences). These genomes showed editing patterns similar to the mouse genome (Figure 1), suggesting that the differences seen between human and mouse stem from unique primate- (or human-) specific factors.

Editing levels vary between different tissues[13,22]. Thus, differences between the tissue distributions of available human and mouse RNA sequences could lead to a bias in the above comparison. To rule out this possibility, we repeated the human-mouse comparison for RNA sequences of the same homogeneous tissue origin. We used three different tissues that have a significant and similar number of sequences for both organisms: brain, thymus and testis. We found (Supplementary tables S1-S3) that for all three tissues the level of editing in human is significant (at least 3% of sequences are edited), while in mouse editing is undetectable for such small datasets. Notably, the editing level in RNAs that originate from the human thymus is exceptionally high: ~17% of sequences and 1:1000bp are apparently edited. Taken together, our results strongly suggest that A-to-I editing patterns significantly differ between human and other organisms.

## RNA editing and the primate-specific Alu element

The vast majority of A-to-I editing detected by our algorithm in human occurs within Alu elements, which are the most abundant SINEs in primates and are frequently contained within transcripts. Although the total number of SINEs in the human and rodents genome is similar [23], forming dsRNA from two consecutive and oppositely oriented SINEs in human is more probable than in mouse since only one SINE is dominant in human. Furthermore, the dsRNAs formed in human are longer (thus contain more adenosines to be edited) since Alu is longer than the equivalent rodent B1 (see supplementary material). An additional possibility is that the Alu repeat could be preferentially targeted by ADARs[13]. We thus suggest that the introduction of Alu elements into the ancestral primate genome is responsible for the large differences in editing patterns between human and other genomes. In addition, the observed difference in editing activity in humans could be related to a human specific splice variant of ADAR2[24,25]. Intriguingly, this splice variant, which accounts for 40% of all ADAR2 transcripts in humans, was created by a birth of Alu-derived exon in intron 7 of ADAR2[16,24].

## Possible implications of abundant RNA editing in human

The vast majority of RNA editing events occurs within non-coding regions of the mRNA. Any role of inosines within these regions is yet a mystery. Possibly, editing of these non-coding regions is meaningless. However, it was already suggested that RNA editing can regulate the triggering of RNA interference (RNAi) and cause or prevent the degradation of the RNA by stabilizing or destabilizing dsRNA stems[26]. Thus, editing in Alu sequences within UTRs might add a powerful mechanism to regulate RNA turnover in primates. In addition, RNA editing was suggested to affect RNA stability[12], localisation[27], and its translation rate[12]. Moreover, editing at the vicinity of a splice site may affect the splicing pattern. In particular, editing can create a new splice site, thus enabling the introduction of new exons[28]. Hence, while the primary role of editing in the UTRs is yet to be revealed, it is already clear that its widespread occurrence in the human transciptome provides evolution with additional, post-transcriptional means allowing fine-tuning of gene expression at the cellular and organism level. In this work we show that abundant editing is unique to primates, and results from the properties of the Alu element repeat. This finding, accompanied by the observations that A-to-I editing is abundant in brain tissues[13,22] and aberrant in a number of neurological disorders[2,7-10], makes it tempting to speculate that widespread editing, as a result of the introduction of Alu elements, may have played a role in the evolution of primates.

### Acknowledgements
The authors thank Compugen's LEADS team for technical assistance and Alon Amit and Michael F. Jantsch for critical reading of the manuscript. The work of E.Y.L. was performed in partial fulfillment of the requirements for a Ph.D. degree from the Sackler Faculty of Medicine, Tel Aviv University, Israel.

## Figures and Tables:

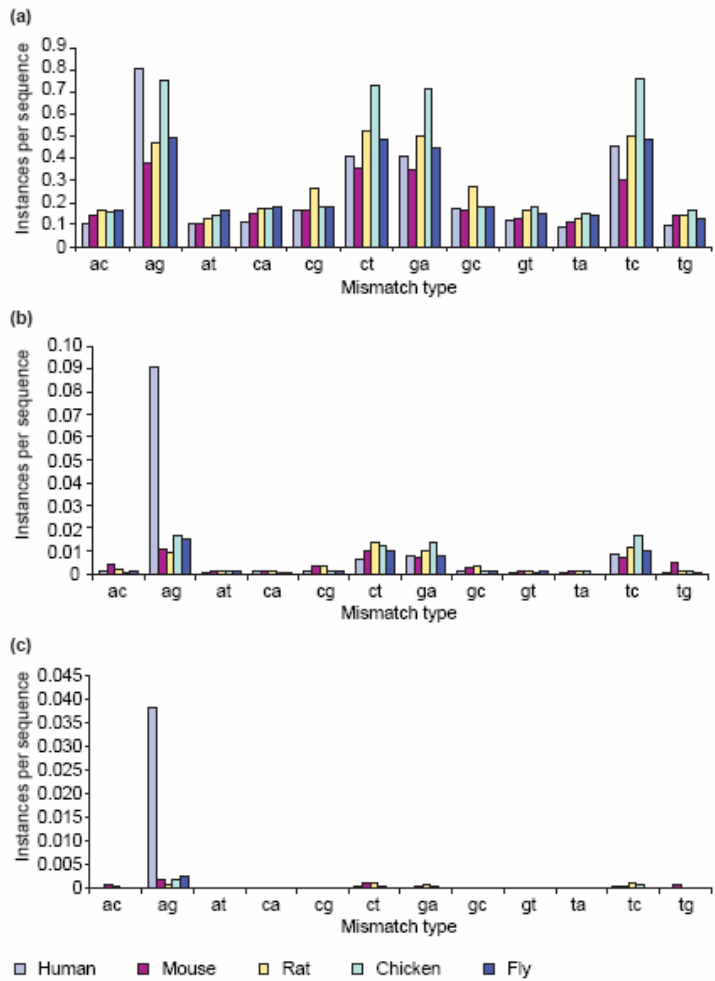

**(d)**

```
GAGGTTGCAGTGAGCCAAGATTATGCCATTGCACTCCAGCCTGGGTGACAAAGCAAGACTCCATCTC                                    Genome
GAGGTTGCAGTGAGCC AGATTATGCCATTGCACTCCAGCCTGGGTGACA AGC  GACTCC TCTC     AY358340
GAGGTTGCAGTGAGCC GGATTATGCCATTGCACTCC GCCTGGGTGACA AGC  AGACTCCATCTC     AI001922
GAGGTTGCAGTGAGCC AGATTATGCCATTGCACTCCAGCCTGGGTGACAAAGCAAGACTCCATCTC     BU740253
GAGGTTGCAGTGAGCCA GATT TGCCATTGCACTCCAGCCTGGGTGACA AGCA GACTCCATCTC      AW614786
GAGGTTGCAGTGAGCCAAGATTATGCCATTGCACTCCAGCCTGGGTGACA AGCAAGACTCC TCTC     BM680006
```

**Figure 1: Mismatches distributions: a multi-species comparison.**
Comparison of the average number of mismatches per sequence for human and mouse RNA sequences shows a significant overabundance of A-to-G mismatches, most probably representing A-to-I editing, in human but not in the other four organisms checked. The distribution of other types of mismatches is similar for all organisms checked. Note that the background mismatch rate in chicken RNAs is significantly higher, presumably due to the lower quality of its current genome draft. X-axis: types of mismatch. The label xy refers to x in the DNA sequence and y in the expressed sequence (e.g., ac label refers to genomic A's that read as C's in the expressed data). Y-axis: the average number of instances of consecutive 1,3 or 5 mismatches per sequence. (A) single mismatches. (B) instances of consecutive three identical mismatches per sequence (C) instances of consecutive five identical mismatches per sequence. (D) Five consecutive editing sites in the 3' UTR of the human ASAM gene. The RNA sequence AY358340 is compared to the genomic sequence. ESTs supporting this editing site are presented. There are 2513 such human mRNAs with 5 consecutive A-to-G mismatches (and only 44 with 5 consecutive G-to-A's or more), comparing to only 127 such mouse mRNAs (57 for G-to-A mismatches).

### Table 1: Mismatches in genome-aligned RNAs

| | Human | | | Mouse | | |
|---|---|---|---|---|---|---|
| | Number of consecutive mismatch | | | Number of consecutive mismatch | | |
| Mismatch | 1 | 3 | 5 | 1 | 3 | 5 |
| A-to-G | 102832(43965) | 11613(4864) | 4926(2513) | 38910(22010) | 1112(901) | 181(127) |
| G-to-A | 52488(34180) | 968(914) | 48(44) | 35876(20226) | 748(681) | 64(57) |
| C-to-T | 52195(34449) | 853(774) | 74(70) | 36750(20116) | 1033(915) | 98(92) |
| T-to-C | 58083(37854) | 1115(1045) | 71(57) | 31630(17207) | 742(681) | 49(46) |
| Fraction of A-to-G instances | 26.5% | 75.4% | 95.0% | 15.1% | 20.2% | 28.0% |

Number of instances of single (or stretches of consecutive) mismatches, for the most common mismatches. The numbers in parentheses are the number of distinct RNA sequences in which the associated instances occur. The last row presents the fraction of A-to-G instances among the total number of all 12 possible mismatches.

# Supplementary material

## *1. Calculating the estimated accuracy*

The overrepresentation of A-to-G mismatches in the various analyses conducted was considered a signature of editing. To obtain an estimate for the number of mismatches resulting from sources other than editing we used the count of G-to-A mismatches, which are at least as common as A-to-G mismatches in sequencing errors, mutations and SNPs, which are the most significant sources of mismatches. Throughout the paper, the accuracy of a set of editing sites is defined as the fraction of the sites or sequences found in excess of the background level. The background level is estimated by the number of sites or sequences found when replacing A-to-G by G-to-A. This estimation of the accuracy level was experimentally confirmed(1) .

## 2. Comparison of editing levels in specific tissues

Editing levels vary between different tissues(1, 2). Thus, the differences between the editing level observed in human and mouse could have resulted from differences in the tissue distributions of available human and mouse RNA sequences. In order to rule out this possibility, we repeated the human-mouse comparison for RNA sequences of homogeneous tissue origin. We used three different tissues that have a significant and similar number of sequences for both organisms: brain, thymus and testis. We found (supplementary tables S1-S3) that for all three tissues the level of editing in human is significant (at least 3% of sequences are edited), while in mouse editing is undetectable for such small datasets. Notably, the editing level in RNAs that originate from the human thymus is exceptionally high: ~17% of sequences and 1:1000bp are edited.

Tables S1-S3: Number of instances of single (or stretches of consecutive) mismatches in RNAs of a specific tissue origin, for the most common mismatches. The numbers in parentheses are the number of distinct RNA sequences in which the associated instances occur. The last row presents the fraction of A-to-G instances among the total number of all 12 mismatches.

### Table S1: brain RNAs

| | Human (5398 sequences) | | | Mouse (1242 sequences) | | |
|---|---|---|---|---|---|---|
| | Number of consecutive mismatch | | | Number of consecutive mismatch | | |
| Mismatch | 1 | 3 | 5 | 1 | 3 | 5 |
| ag | 6469(2298) | 974(391) | 431(219) | 756(353) | 19(19) | 2(2) |
| ga | 2485(1698) | 38(37) | 1(1) | 707(321) | 19(16) | 1(1) |
| ct | 2652(1806) | 43(37) | 5(5) | 792(366) | 21(21) | 3(3) |
| Tc | 3274(2108) | 74(66) | 5(4) | 768(334) | 17(16) | 1(1) |
| Fraction of A-to-G instances | 33.8% | 85.1% | 97.1% | 16.2% | 20.4% | 25.0% |

## Table S2: thymus RNAs

| | Human (1090 sequences) | | | Mouse (3746 sequences) | | |
|---|---|---|---|---|---|---|
| | Number of consecutive mismatch | | | Number of consecutive mismatch | | |
| Mismatch | 1 | 3 | 5 | 1 | 3 | 5 |
| ag | 3036(670) | 633(193) | 304(134) | 779(554) | 28(18) | 8(5) |
| ga | 573(374) | 9(9) | 0(0) | 783(575) | 10(10) | 1(1) |
| ct | 601(394) | 7(7) | 0(0) | 561(371) | 7(7) | 0(0) |
| tc | 810(511) | 18(11) | 5(1) | 455(310) | 8(8) | 0(0) |
| Fraction of A-to-G instances | 50.5% | 94.5% | 98.1% | 14.4% | 26.9% | 72.7% |

## Table S3: testis RNAs

| | Human (6217 sequences) | | | Mouse (6881 sequences) | | |
|---|---|---|---|---|---|---|
| | Number of consecutive mismatch | | | Number of consecutive mismatch | | |
| Mismatch | 1 | 3 | 5 | 1 | 3 | 5 |
| ag | 5142(2674) | 394(209) | 133(64) | 1717(1180) | 32(28) | 2(2) |
| ga | 2769(1918) | 39(37) | 4(3) | 1707(1079) | 49(42) | 8(7) |
| ct | 2674(1894) | 27(26) | 0(0) | 1699(1040) | 52(46) | 6(5) |
| tc | 3037(2024) | 44(43) | 0(0) | 1148(729) | 30(27) | 3(3) |
| Fraction of A-to-G instances | 27.6% | 75.9% | 95.7% | 12.6% | 11.7% | 5.9% |

## 3. Alu and the number of potential dsRNAs

The vast majority of A-to-I editing detectable through our algorithm in human occur within Alu elements, which are the most abundant SINEs in primates. Alu elements tend to accumulate within genes(3), and are present in about 75% of all human genes(4). Thus, Alu repeats occurring within the RNA transcript facilitates the formation of the dsRNA substrates required for ADARs action by pairing with other Alu repeats within the pre-mRNA.

It is surprising to find such a substantial difference in global A-to-I editing patterns between human and rodents, because the number of SINEs in the human genome is similar to the total number of rodent SINEs(5). One major reason for the difference is the fact that only one SINE is dominant in human, making a dsRNA formation out of two consecutive and oppositely oriented SINEs more probable. Furthermore, the dsRNAs formed in human are longer (thus contain more adenosines to be edited) since Alu is longer than the equivalent rodent B1. We checked whether this effect alone can explain the ~20 fold difference in the number of editing sites. For this purpose, we took all human ESTs and cDNAs and aligned them to the genome (details of this procedure are given in Sorek et al(6)). Following Levanon et al(1), we aligned the expressed part of the gene with the corresponding genomic region, looking for reverse complement alignments longer than 32nt with identity levels higher than 85%. We found 429,000 such potential dsRNAs structures, covering 4.69Mbp. In comparison, the number of such potential dsRNAs structures in the mouse genome was 81,000, covering 969kbp. Thus, the number of potential dsRNA stems in human is only 5-fold larger than that of mouse, and similarly, the total size of expressed DNA regions potentially creating such stems is only 5 times larger, comparing to the 20-to-40 fold increase in editing observed in human. It is thus possible that the Alu repeat, besides being more abundant than each of the mouse repeats, is also preferentially targeted by ADARs. In fact, we have previously reported that Alu elements contain "hot-spots" for editing(1).

## 4. Comparison of editing levels in sequences of a single source

Another potential bias that could have led to the differences between the editing levels observed in human and mouse might result from the fact that different sources are used for RNA purification and sequencing in different research centres. In order to rule out this possibility, we repeated the above analysis for RNA sequences of homogeneous origin. We scanned more than 22,000 full-length human and mouse cDNA sequences coming from the Mammalian Gene Collection (MGC) Program(7), and repeated the analysis for these sequences only. Comparing the number of sequences with 3 consecutive A-to-G mismatches to the corresponding number with G-to-A mismatches we find that ~180 human sequences were edited. In contrast, editing is undetectable in the mouse data.

## 5. Library distribution of edited RNAs

To make sure the overrepresentation of edited RNAs in human is not a result of a small number of faulty libraries, we have studied the library assignment of human RNAs that exhibit editing. In particular, we studied the 2513 RNA sequences that show five consecutive A-to-G mismatches. Library assignment is available for 1832 of these sequences. We find that they come from 207 different libraries. In addition, we found that at least one sequence with 5 or more consecutive A-to-G mismatches is present in 170 out of the 230 libraries (74%) containing at least 50 RNA sequences.

## *6. Materials and methods*

Employing the algorithm of Ref. 11, human and mouse ESTs and cDNAs were obtained from NCBI GenBank version 136 (June 2003; www.ncbi.nlm.nih.gov/dbEST). The genomic sequences were taken from the human genome build 33, June 2003, and mouse genome build 32, November 2003, (data downloadable from www.ncbi.nlm.nih.gov/genome/guide/human).

Sequences were cleaned from terminal vector sequences, and low-complexity stretches and repeats (including Alu repeats) in the expressed sequences were masked. Then, expressed sequences were compared with the genome to find likely high-quality hits. They were then aligned to the genome by use of a spliced alignment model that allows long gaps. Only sequences having >94% identity to a stretch in the genome were used in further stages. Further details can be found in Sorek at al(6).

In addition, RNA sequences and their pair-wise alignments to the genome were downloaded from UCSC genome browser site http://genome.ucsc.edu (human assembly July 2003, mouse assembly Oct. 2003, rat assembly June 2003, chicken assembly Feb. 2004 and fly assembly Jan. 2003). We kept only sequences with a unique alignment to the genome, and then recorded all mismatches.